\documentclass[
reprint,
groupedaddress,
showpacs,
showkeys,
amsmath, amssymb,
aps,
prl,
]{revtex4-1}

\usepackage{xcolor}
\usepackage{graphicx}% Include figure files
\usepackage{dcolumn}% Align table columns on decimal point
\usepackage{bm}% bold math
\usepackage[colorlinks=true, allcolors=blue]{hyperref}% add 
\usepackage{comment}
\graphicspath{{figures/}}
\usepackage[separate-uncertainty=false]{siunitx}
\usepackage{setspace}

\usepackage{xspace}

\newcommand{\Tr}{\mathrm{Tr}}
\newcommand{\D}{\mathrm{div}}
\newcommand{\V}[1]{\mathbf{#1}}

\usepackage{xstring}
\newcommand*{\aref}[1]{%
	\IfBeginWith{#1}{eq:}{Eq.~\eqref{#1}}{}%r
	\IfBeginWith{#1}{fig:}{Fig.~\ref{#1}}{}%
	\IfBeginWith{#1}{tab:}{Table~\ref{#1}}{}%
	\IfBeginWith{#1}{appendix:}{Appendix~\ref{#1}}{}%
	\IfBeginWith{#1}{sec:}{Section~\ref{#1}}{}%
}

\newcommand\varpm{\mathbin{\vcenter{\hbox{%
 \oalign{\hfil$\scriptstyle+$\hfil\cr
     \noalign{\kern-.3ex}
     $\scriptscriptstyle({-})$\cr}%
}}}}

\begin{document}

\title{
Quantum-torque-induced breaking of magnetic interfaces in ultracold gases}
\date{\today}

\author{A. Farolfi}
\author{A. Zenesini$^*$}
\author{D. Trypogeorgos$^\dagger$}
\author{C. Mordini$^\ddagger$}
\author{A. Gallem\'i}
\author{A. Roy}
\author{A. Recati$^*$}
\author{G. Lamporesi$^*$}
\author{G. Ferrari}

\affiliation{INO-CNR BEC Center and Dipartimento di Fisica, Universit\`a di Trento, and Trento Institute for Fundamental Physics and Applications, INFN, 38123 Povo, Italy. }

\maketitle

%Introductory paragraph

\textbf{A rich variety of physical effects in spin dynamics arises at the interface between different magnetic materials \cite{Zabel2008}. Engineered systems with interlaced magnetic structures have been used to implement spin transistors, memories and other spintronic devices \cite{Wolf2001,Locatelli2014}. However, experiments in solid state systems can be difficult to interpret because of disorder and losses. Here, we realize analogues of magnetic junctions using a coherently-coupled mixture of ultracold bosonic gases. The spatial inhomogeneity of the atomic gas makes the system change its behavior from regions with oscillating magnetization --- resembling a magnetic material in the presence of an external transverse field --- to regions with a defined magnetization, as in magnetic materials with a ferromagnetic anisotropy stronger than external fields. Starting from a far-from-equilibrium fully polarized state, magnetic interfaces rapidly form. At the interfaces, we observe the formation of short-wavelength magnetic waves. They are generated by a quantum torque contribution to the spin current and produce strong spatial anticorrelations in the magnetization. Our results establish ultracold gases as a platform for the study of far-from-equilibrium spin dynamics in regimes that are not easily accessible in solid-state systems.}

\bigskip

%Main Text

The local magnetization in a magnetic material evolves depending on three ingredients: the external magnetic field, the nonlinear ferromagnetic anisotropy and the inhomogeneity of the magnetization itself. The evolution can be described by using the well known Landau-Lifshitz equation (LLE) \cite{Landau1935,Baryakhtar2015}. When a large external magnetic field is applied, all spins precess around it. 
For a vanishing field amplitude, the spins precess around a preferential spatial direction characteristic of the material itself, given by the magnetic anisotropy. In real materials, this anisotropy is usually very small compared to the lowest technically achievable uniform external field \cite{Baryakhtar2015}. The term of the LLE incorporating the inhomogeneity -- derivable from the Heisenberg exchange term -- becomes particularly relevant in the presence of magnetic interfaces. In the absence of such an exchange term, the LLE reduces to the Josephson equations for Bose-Einstein condensates \cite{Josephson1962, Smerzi1997}. Similarly to a magnet in an external field, Josephson equations have different dynamical regimes: either the system oscillates between two states (precession around external field) or it is self-trapped in one of them (precession around the dominant anisotropy direction).

\begin{figure}
%  \centering
  \includegraphics[width=1.\columnwidth]{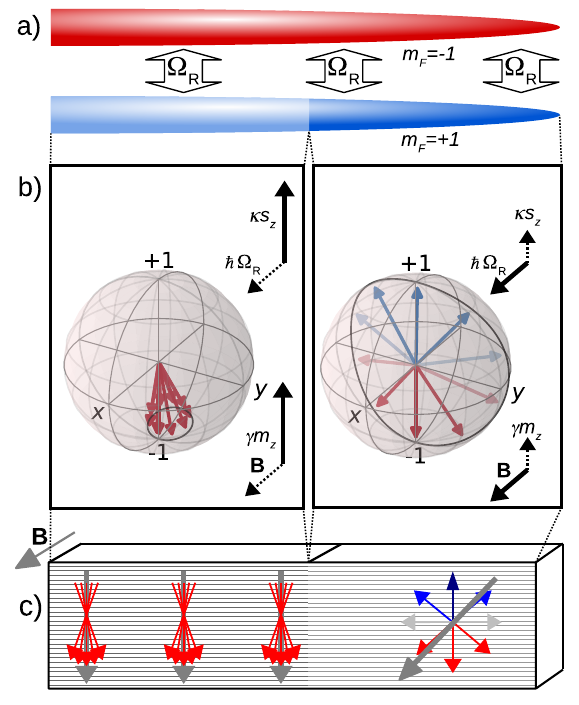}  
  \caption{ \textbf{Analogy between a coherently-coupled atomic mixture and a magnetic heterostructure.} \textbf{a}, Sketch of the trapped ultracold atomic mixture in the two hyperfine states $|1,+1\rangle$ (blue) and $|1,-1\rangle$ (red), coupled via coherent radiation with strength $\Omega_\mathrm{R}$. \textbf{b}, Local evolution of the system, represented on the Bloch sphere in the center (where interactions $\kappa s_z$ exceed the coupling $\Omega_\mathrm{R}$) and in the tails (where the coupling exceeds the interaction term). \textbf{c}, Pictorial view of the magnetic analogue. The material has a spatially varying ferromagnetic anisotropy $\gamma$, smaller or larger than the external magnetic field \textbf{B}, respectively on the external or internal regions. The equilibrium magnetization of the material (grey arrow) follows the dominating effect between intrinsic anisotropy and external field. Magnetic interfaces are present between these regions. Black arrows in panel \textbf{b} represent the contributions of the different physical quantities both for the atomic system (top) and for its magnetic analogue (bottom).  }
  \label{fig:1}
\end{figure}

In our experiment, we coherently couple two hyperfine states, $|F,m_F\rangle=|1,\pm1\rangle$, of a near-zero-temperature Bose-Einstein condensate of $^{23}$Na atoms, trapped in an elongated harmonic potential (see Methods); $F$ is the total atomic angular momentum and $m_\text{F}$ is its projection along the quantization axis  $z$. This system is equivalent to a magnetic material with nonuniform magnetic anisotropy in the presence of an external transverse field, as sketched in \aref{fig:1}. 
In the analogy, the effective external magnetic field is represented by the electromagnetic radiation that coherently couples the spin states, while the ferromagnetic anisotropy is due to the nonsymmetric interatomic interactions and varies spatially thanks to the density inhomogeneity of the sample.
The interatomic interaction energy of sodium mixtures, together with the low magnetic field noise of our magnetically shielded system, allows us to tune the coupling strength below the interaction energy. This corresponds, in magnetic materials, to an external field below the magnetic anisotropy.
If all spins are initially aligned along $z$ and a field is suddenly applied, they start precessing, as illustrated by the vector on the Bloch sphere (\aref{fig:1}b). The precession happens around a direction that depends on the local properties. While standard magnetic materials quickly align along one axis (grey vector in \aref{fig:1}c) because of dissipation, our nondissipative atomic system allows to study a longer dynamical evolution. 

The atomic gas can be described with a two-component order parameter $\Psi(x)=(\psi_{+1},\psi_{-1})^T$, where $\psi_\alpha$ is the macroscopic wave function of the Bose-Einstein condensate in the state $\alpha=\pm 1$. The tight confinement along two spatial directions makes the spatial dynamics be essentially one-dimensional along the $x$ direction (see Methods). Therefore, the state of the system is fully described by the density matrix $(\Psi^*\otimes \Psi)(x)=\{\psi_\alpha^*(x)\psi_\beta(x)\}_{\alpha,\beta=\pm 1}$. The density matrix is composed of a scalar part, $n=\Tr(\Psi^*\otimes \Psi)$, corresponding to the total density of the condensate, and of the spin-density $ \mathbf{s}=\Tr(\boldsymbol{\sigma}\Psi^*\otimes \Psi)$, with $|\mathbf{s}|=n$ and $\boldsymbol{\sigma}$ representing the vector of Pauli matrices. Hereafter vector quantities are defined on the Bloch sphere.

In general the dynamics is described (see Methods) by coupled differential equations for $n$, $ \mathbf{s}$ and the velocity field $v=j/n$, where $j$ is the atom density current \cite{Nikuni2003}. 
Since the total atom number is a conserved quantity, $n$ satisfies the continuity equation, with the purely advective current $j$: $\dot{n}+\partial_x j$=0. The equation of motion of $\mathbf{s}$ reflects the possibility of twisting the spin and the absence of spin conservation. Both features are due to the combination of the coherent Rabi coupling and the lack of SU(2) symmetry of the non-driven system. The Rabi coupling is described by the linear transverse field $\Omega_\mathrm{R}\hat{x}$. The lack of SU(2) symmetry leads to a nonlinear field $\kappa s_z\hat{z}$, with $\kappa$ proportional to the difference between intra- and intercomponent interactions, $\delta g$, and including the effect of the dimensional reduction (see Methods). The spin equation of motion can be written as 
\begin{equation}
  \dot{\mathbf{s}}+\partial_x \mathbf{j}_s =\mathbf{H(\mathbf{s})}\times\mathbf{s},
\label{spindyn}
\end{equation}
where we introduce the effective magnetic field $\mathbf{H}= \Omega_\mathrm{R}\hat{x}+\kappa s_z\hat{z}$. The spin current reads
\begin{equation}\label{eq:j}
  \mathbf{j}_s=v\mathbf{s}+\frac{\hbar}{2mn}\partial_x \mathbf{s}\times\mathbf{s}.
\end{equation}
The first term is the spin advection. The spatial derivative of the second one is the \textit{quantum torque}, which depends explicitly on $\hbar/m$. Remarkably, the quantum torque originates as a pure quantum effect, vanishing when $\hbar$ is set to zero, or equivalently when the mass of the atoms is infinite (see classical analogue in Ref.\cite{Pigneur2018class}).

The equation of motion for the spin density, Eq.\,(\ref{spindyn}), in the absence of spin advection, is equivalent (see Methods) to a non-dissipative LLE  \cite{Nikuni2003}. 
Therefore, if the density and velocity dynamics can be neglected, the dynamics of a coherently-coupled Bose gas mimics the magnetization dynamics in a magnetic sample, where the quantum torque plays the role of the exchange term. 
Since the quantum torque depends on the curvature of $\mathbf{s}$, it plays a crucial role in the presence of magnetic interfaces. Often in literature the effective field in the LLE includes the torque as well \cite{Baryakhtar2015}, which in the magnetic context is due to the exchange interaction.

\begin{figure}
 \includegraphics[width=1.\columnwidth]{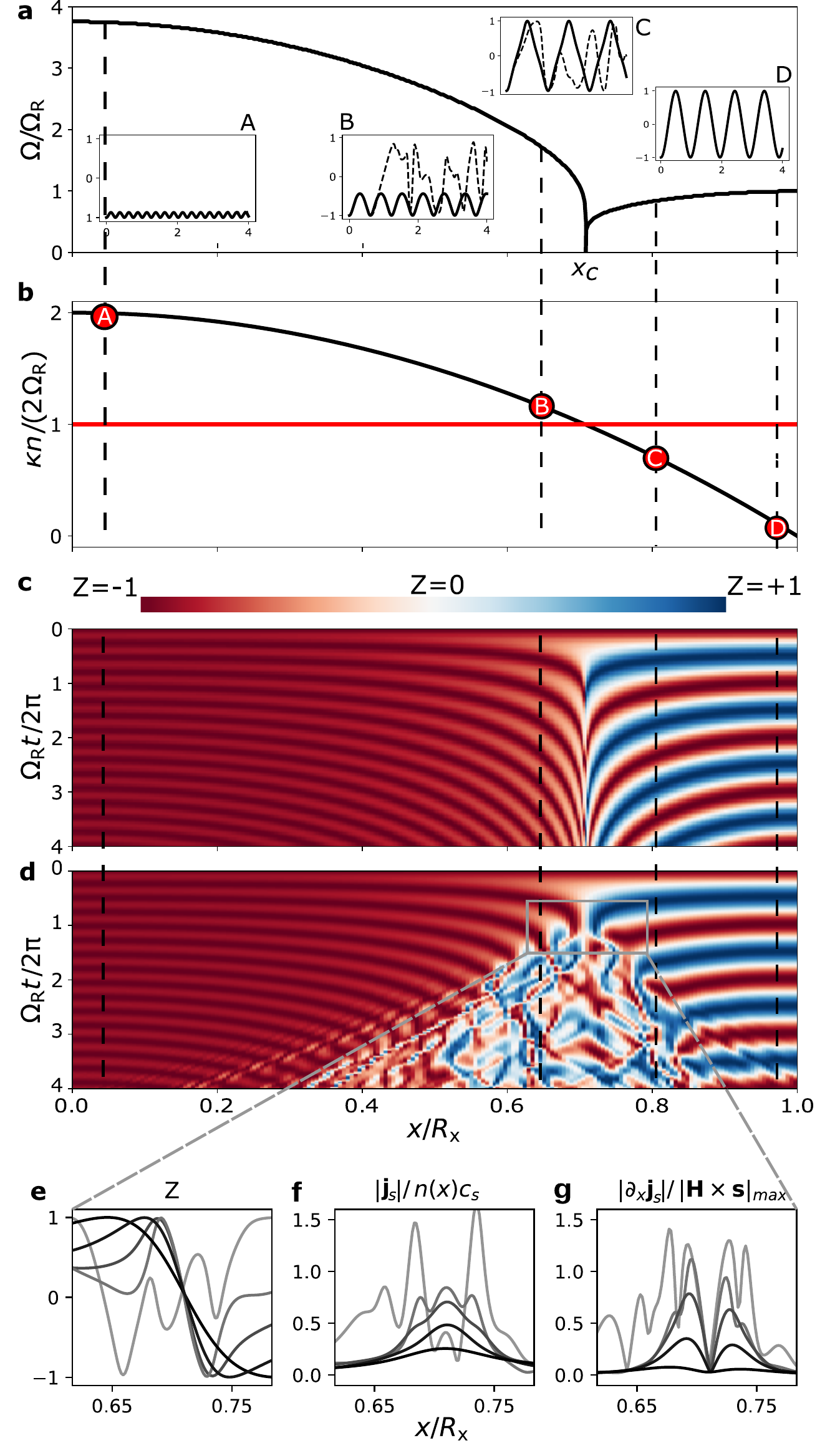}
  \caption{ \textbf{Quantum torque effect at the interface.} \textbf{a}, Main local oscillation frequency $\Omega$ of the relative magnetization $Z$ as a function of the position in the gas. The softening of the precession frequency at $x_c$ marks the transition between the two dynamical regimes. The insets show the time evolution of $Z(t)$ for four different spatial points A-D (dashed lines). \textbf{b}, Spatial variation of the magnetic anisotropy. The red line marks the critical value. 
  \textbf{c-d}, Time evolution of the magnetization across the cloud according to LLE without (\textbf{c}) and with (\textbf{d}) quantum torque contribution. 
  Four different spatial points are considered: deep in the self-trapped regime (A), weakly self-trapped (B), weakly oscillating regime (C) and deep in the oscillating regime (D).
  Continuous and dashed lines in the inset of \textbf{a} correspond to the time evolution without and with the quantum torque term, respectively. Note that in B and C, the local dynamics is disrupted due to the excitation of short wavelength magnetic waves.
  \textbf{e-g}, Profiles of $Z$, $|\mathbf{j}_s|/nc_s$, with $2m c_s^2=\hbar\kappa n$, and $|\partial_x \mathbf{j}_s| / \, |\mathbf{H} \times \mathbf{s}|_{max}$ calculated for $\Omega_{\mathrm{R}}t / 2 \pi=$0.7,\,0.8,\,0.9,\,1,\,1.5 around the interface breaking (black to gray).
    }
  \label{fig:2}
\end{figure}

Taking advantage of the absence of dissipative terms in Eq.~(\ref{spindyn}), we study the long time dynamics of systems with far-from-equilibrium initial configurations. 
Before discussing the actual experimental configuration (\aref{fig:1}a), it is useful to consider the dynamics in the simple spatially homogeneous case. Equation~(\ref{spindyn}) reduces to $ \dot{\mathbf{s}} =\mathbf{H(\mathbf{s})}\times\mathbf{s}$, i.e., equivalent to the Josephson equations for weakly-interacting Bose gases \cite{Smerzi1997}. These equations are usually written in terms of the relative magnetization $Z=s_z/n$ and the relative phase $\phi=\arctan(s_y/s_x)$, which in \aref{fig:1}b correspond to the projection of the quantum state on the $z$ axis of the Bloch sphere, and its equatorial angle, respectively, $\mathbf{s}=n(\sqrt{1-Z^2}\cos\phi,\sqrt{1-Z^2}\sin\phi,Z)$. 

With an initially fully polarized state $s_z=- n$, two different dynamical regimes are possible: (i) for $\Omega_\mathrm{R}>|\kappa n/2 |$, the magnetization oscillates between $s_z=\pm n$ with a frequency $\Omega$ (see the dynamics in C and D in \aref{fig:2}b and associated continuous line in insets), a.k.a. Josephson oscillations; (ii) for $\Omega_\mathrm{R}<|\kappa n/2 |$, the system enters the so-called self-trapped regime \cite{Smerzi1997,Zibold2010}, where the spin precesses such that $-n \le s_z(t)\le \text{max}(s_z)<0$. In the self-trapped regime, $Z$ never changes sign (see the dynamics in A and B in \aref{fig:2}b).
Interestingly, the precession frequency -- not only the amplitude -- drastically changes across the transition, as shown in \aref{fig:2}a, with a softening of the precession frequency at the transition point.

A harmonically trapped Bose-Einstein condensate in the Thomas-Fermi approximation shows a density reduction from the center outwards, $n(x)=n_0(1-x^2/R_\text{x}^2)$ (black line in \aref{fig:2}b).
Thanks to this inhomogeneity, we can realize systems that present simultaneously both behaviors in spatially different regions.
If the central density $n_0$ is large enough for the system to be locally in the self-trapped regime, there exists a position $x_c$ that separates this region from the low-density one, where Rabi-like oscillations of the magnetization occur. The sharp transition from one regime to the other creates an interface.

When the system is prepared in an initially fully polarised state $|1,-1\rangle$, after about $1/\Omega_\mathrm{R}$, the formation of magnetic interfaces takes place at $x_c$. In the absence of spin current, the interfaces would stand between the two dynamically different regions, which would follow standard Josephson dynamics. 
However, quantum torque becomes relevant at the interfaces and breaks the Josephson dynamics. Figures~\ref{fig:2}c and \ref{fig:2}d show the theoretical evolution of the magnetization throughout the sample, respectively without and with the quantum torque term.

Figure~\ref{fig:2}e-g compare the profiles of $Z$, $\mathbf{j}_s$ and $\partial_x \mathbf{j}_s$ near the interface for 5 consecutive times (from black to gray). They clearly illustrate that, while the magnetization gradient rises at the interface, the spin current rapidly increases and its spatial derivative peaks. When the quantum torque term 
becomes of the order of $|\mathbf{H} \times \mathbf{s}|$, it counteracts the magnetization gradient forming 
magnetic waves with short wavelength.

\begin{figure*}
  \includegraphics[width=1.9\columnwidth]{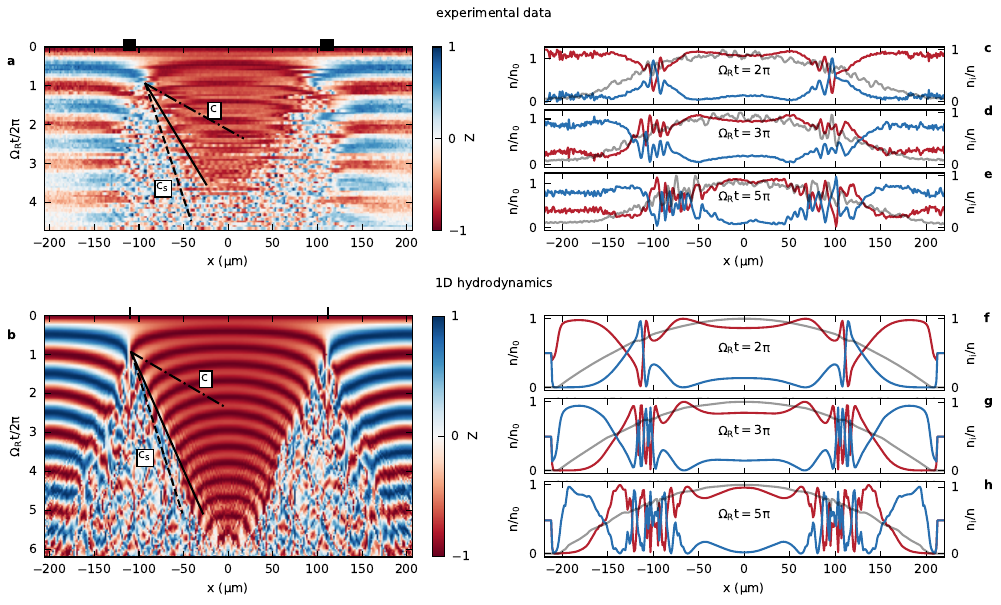}
  \caption{\textbf{Evolution of the magnetization.}
  \textbf{a}, Measured magnetization along the non-uniform 1D gas as a function of time in the presence of a small coupling ($\Omega_\mathrm{R}=2\pi \times118 $ Hz$\,= 0.33\, |\kappa n_0|$).
  Panel \textbf{b} shows the full 1D hydrodynamic simulation of the same system.
  At the interface, around the critical position $x_c$ (black bars), we observe a net change in the behavior from the oscillating to the self-trapped regime. The width of the bars is evaluated considering the uncertainties in $n$ and $R_x$. Between the two regions, a new one, characterized by a strong fluctuation of the magnetization, emerges and grows. The fluctuating region expands in the self-trapped region with a constant velocity (continuous line),
  estimated to be $v=1.8(2) c_s$ (experimental data) and $v=1.4(1) c_s$ (numerical data).
  The dashed and dot-dashed lines show the slope corresponding to the local spin sound $c_s$ and sound velocity $c$.
  \textbf{c-e}, Measured density profile of each component (red, blue), normalized to the local total density (gray), for $\Omega_\text{R}t=2\pi$, $3\pi$ and $5\pi$. \textbf{f-h}, Same as \textbf{c-e}, but evaluated numerically. Notice that the experimental profiles are taken after a short expansion (see Methods) that enhances the total density modulation.}
  \label{fig:3}
\end{figure*}

Other systems of atomic mixtures have been studied in this context, but none observed both regions in a single sample and the sharp interface separating them.
Either self-trapped or oscillating regimes were observed in "zero"-dimensional (single mode) systems both with Rabi- \cite{Zibold2010} and tunnel-coupling \cite{Albiez2005}. 
Complex dynamics was observed in elongated, inhomogeneous BECs in the presence of Rabi coupling \cite{Matthews1999,Nicklas2011}. Double-well potentials were also used to investigate Josephson dynamics both in bosonic \cite{LeBlanc2011, Spagnolli2017, Pigneur2018, Mennemann2020} and fermionic \cite{Luick2019, Kwon2020} gases.

Figure~\ref{fig:3}a shows the experimental measurement of the magnetization. 
We let the system evolve in the presence of coherent coupling with $\Omega_\mathrm{R}=0.33\,|\kappa n_0|$ for a variable time $t$ and we separately image the two spin populations. This allows us to extract the local magnetization (see Methods). For each experimental run, we integrate the magnetization of the elongated atomic sample in the radial directions and obtain $Z(x)$.
Combining the measurements of $Z(x)$ at different times $t$, we reconstruct the full dynamics.
As predicted by theory, the system spatially explores two completely different regimes, depending on the ratio between the driving frequency $\Omega_\mathrm{R}$ and $|\kappa n(x)|$.
At short time, we observe the creation of two magnetic interfaces at 
positions $\pm x_c$, where the condition $ \Omega_\mathrm{R}=\kappa n(x_c)/2$ is matched.
The magnetization in the central region slightly oscillates, never changing sign, while in the outer part atoms undergo full oscillations at a frequency close to $\Omega_\mathrm{R}$.
As time goes on, the interfaces break, the self-trapped region becomes smaller and smaller and strongly fluctuating regions are created and grow in size. 
In Fig 3b we report the corresponding results obtained with the full hydrodynamic simulation, that we numerically implement by solving the equation of motion for $\Psi$, represented by two coupled Gross-Pitaevskii equations \cite{Pitaevskii16}.
Figure~\ref{fig:3}f-h and ExtendedDataFig.1  
show that the density and the superfluid velocity terms are irrelevant for the dynamics. Therefore for our protocol, the LLE provides a very good description of the system dynamics.
We impute the difference between experimental data and simulation to a possible mismatch between the actual parameters in the experiment and those used in the simulations.
Deep in both the oscillating and self-trapped regions one can see a smooth spatial variation. In the strongly fluctuating region, instead, the magnetization varies on a very small length scale, as visible also in the measured (\aref{fig:3}c-e) and numerically simulated (\aref{fig:3}f-h) density profiles.

\begin{figure}
  \centering
  \includegraphics[width=1.\columnwidth]{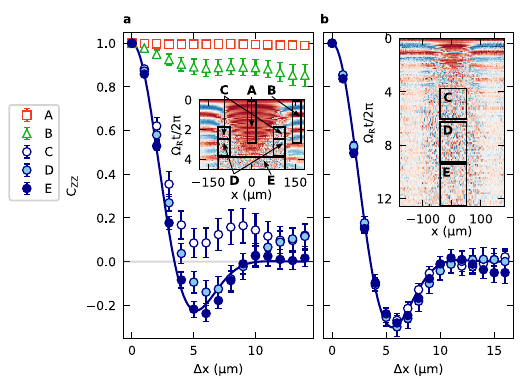}
  \caption{
  \textbf{Correlation of the magnetization. a,}  Spatial autocorrelation of the magnetization for different regions of the sample during the evolution, indicated by the rectangles in the inset. In the self-trapped (A) and oscillating (B) regions, the long-range order of the sample is maintained. In the excited region (C-E), the coherence has a minimum at around 5-6 \SI{}{\um}. At longer times the minimum becomes negative. Error bars corresponds to s.d. of the correlation at different times averaged in the rectangles.  \textbf{b,} Same as \textbf{a}) but for the magnetization reported in the inset, that was obtained from a different dataset ($\Omega=2\pi\times 62$ Hz$\,=0.28|kn_0|$) and extended to longer times.
  }
  \label{fig:4}
\end{figure}

In order to characterize the evolution of the strongly fluctuating region, we estimate the wave front speed as it propagates in the inner region with almost uniform density. Note that the strongly inhomogeneous density and the increasing role of thermal component in the outer regions make it harder to study the wave front propagating outwards and to compare it with the zero-temperature simulations.
We identify the edge of the fluctuating region by calculating the standard deviation of the magnetization in \SI{5}{\um}-wide windows, which exhibits a jump from a low to a nonzero value as the self-trapped regime is consumed by the fluctuating regions.
We observe that the front propagates with a rather constant speed as indicated by the solid line in \aref{fig:3}a.
The propagation speed $v=3.1(1)\,$mm/s (solid) is larger than the local spin sound velocity $c_s\simeq \SI{1.7}{mm/s}$ (dashed), while it remains well below the local density sound velocity $c\simeq\SI{9.2}{mm/s}$ (dot-dashed). We estimate the two velocities from $2m c_s^2=\hbar\kappa n$, and $m c^2=\hbar\kappa n g/(g-g_{12})$, respectively, where $g$ and $g_{12}$ are the interaction constants (see Methods).

Figure~\ref{fig:4}a shows the correlations of the magnetization $C_\text{ZZ}(\Delta x)=\int Z(x)Z(x+\Delta x)/\int Z(x)^2$ far from the breaking point, calculated on the data reported in \aref{fig:3}a.
The integral is performed in the areas enclosed in the rectangles shown in the insets. While in the self-trapped (A) and oscillating (B) regions the system maintains correlations over a long spatial scale, the fluctuating regions (C) show a strong decay.
On longer timescales (D-E), the autocorrelation shows a negative minimum around 5-6 \SI{}{\um}. Its position does not depend on the time in the evolution, nor on the position where correlations are calculated. 
On long time scales the function stabilizes to a well defined shape showing anticorrelations.
Figure \ref{fig:4}b shows the same analysis on a different dataset (inset) extending to longer times, where we do not observe any further evolution of the autocorrelation.

The continuous lines are fits to the data in regions E with the function $f(x)=\cos(\pi x/L_1)\exp(-(x^2/2L_2^2))$. For panel b) the best-fit parameters are $L_1=\SI{7.1(1)}{\um}$ and $L_2=\SI{4.0(1)}{\um}$. Such a behavior of the magnetic correlations indicates a well-defined size of the magnetic domains ($\simeq L_1$) that is larger than the spin healing length $\xi_s=
\hbar / \sqrt{ 2 m n \delta g } \approx \SI{0.8}{\um}$ and the Rabi healing length $\xi_\mathrm{R}=\sqrt{\hbar/m\Omega_\mathrm{R}} \approx \SI{2.6}{\um}$.

The observed size of the magnetic domains and the comparable inward and outward wave front propagation, visible in the simulation, trigger interesting questions about the origin of the spin structures. 
The density profiles (\aref{fig:3}c-h), the fact that the speed of the magnetic wave front is larger than typical Landau spin critical velocity (\aref{fig:4}), and the spin correlations in the downstream, suggest that these excitations are closely related to magnetic shock waves \cite{Kosevich1990,Congy2016,Ivanov2017}. 
Shock waves have been studied in single component ultracold atomic gases \cite{Chang2008,Mossman2018} and very recently, in the presence of spin-orbit coupling \cite{mossman2020}.
However, the LLE in the presence of both transverse magnetic field and anisotropy are not integrable and have been shown to present a chaotic behavior \cite{Daniel92}. Therefore our protocol could excite a new kind of magnetic shock waves with a chaotic character, leading to a turbulent behavior of the magnetization that might have connections with a spin glass \cite{Tsubota2013}.
Even on the pure theoretical side, such kind of waves have never been studied and thus deserve further analysis.

\section{methods}

\subsection{One dimensional superfluid spinor hydrodynamics and LLE}
In this section we briefly explain how the spinor superfluid hydrodynamics can, in certain situations, be reduced to the Landau-Lifshitz equations. 

At zero temperature, a two-component Bose gas composed of atoms of mass $m$ is described by superfluid spinor hydrodynamics, that is equivalent to the Gross-Pitaevskii description, which provides the dynamics of the total density $n$, the superfluid velocity $\V{v}$ and the spin $\V{s}$. The hydrodynamics equations can be written as (see, e.g., \cite{Nikuni2003}): 
\begin{eqnarray}
\label{shydro}
& \dot{n}+\D(n \V{v})=0,\\
&m\dot{\V{v}}+\D\left( \frac{mv^2}{2}+\mu+\frac{s_z}{n}h+V-\frac{\hbar^2\nabla^2\sqrt{n}}{2m\sqrt{n}}+\frac{\hbar^2|\nabla\mathbf{s}|^2}{8m n^2}\right)=0,\\
&\dot{\mathbf{s}}+\sum_{\alpha}\partial_\alpha(\V{j}_{s,\alpha})=\mathbf{H}(\V{s})\times{\V{s}},\label{eq:spin}
\end{eqnarray}
where $\alpha=x,y,z$ indicates here the three spatial directions.

The first equation is the standard continuity equation and  the second one is the Euler equation for the superfluid velocity, where the chemical potential $\mu$ and its spin counterpart $h$ are given by 
\begin{equation}
\label{muandh}
\begin{split}
    \mu[n,s_z]&=\frac{1}{2}(g+g_{12})n+\frac{\hbar\Omega_\mathrm{R}}{2}\frac{n}{n^2-s_z^2}s_x\\
    h[n,s_z]&=\frac{1}{2}(g-g_{12})s_z-\frac{\hbar\Omega_\mathrm{R}}{2}
    \frac{s_z}{n^2-s_z^2}s_x,
\end{split}
\end{equation}
with $g$ and $g_{12}$ the intra- and intercomponent interaction strengths, respectively, and $\Omega_\mathrm{R}$ the Rabi coupling strength. Let us also remind here that for a spinor superfluid, the velocity is generally not irrotational. Equation~(\ref{eq:spin}) is the most important one since it determines the spin dynamics. It shows that the total spin is not a conserved quantity as long as the effective magnetic field $\mathbf{H}(\V{s})=(\Omega_\mathrm{R},0,(g-g_{12})s_z)$ is non-zero. The spin current is a tensor which contains two contributions:
\begin{equation}
    \V{j}_{s,\alpha}=v_\alpha\mathbf{s}+\frac{\hbar}{2m}\left(\partial_\alpha \mathbf{s}\times\frac{\mathbf{s}}{n}\right), 
\end{equation}
where the first term corresponds to the classical spin advection and the second is the current due to the spin-twist and at the origin of the quantum torque term in \aref{eq:spin} discussed in the main text.

If the dynamics is such that the density and the superfluid velocity are constant, the system is described only by \aref{eq:spin}. The latter is formally equivalent to the so-called Landau-Lifshitz equation (LLE) for the magnetisation dynamics in magnetic materials (see, e.g., \cite{Baryakhtar2015}). 

For the geometry of the experiment, we also have that the spin dynamics is effectively one-dimensional, which, formally  corresponds to have the possibility of writing 
\begin{equation}
\V{s}(\V{r},t)=f(x,\V{r_\perp})\V{s}(x,t),\label{1dred}
\end{equation} with $x$ the axial coordinate, $\V{r}_\perp$ the transverse coordinates (irrelevant for the spin dynamics) and $f(x,\V{r}_\perp)=(1-x^2/R_x^2-r_\perp^2/R_r^2)/(\pi R_r^2)$ a function taking into account the Thomas-Fermi density profile, with radii $R_x$ and $R_r$. Substituting the Ansatz (\ref{1dred}) into \aref{eq:spin} and integrating over $\V{r}_\perp$, one obtains the one-dimensional LLE Eq. (2) of the main text with 
\begin{equation}
    \hbar\kappa=(g-g_{12})\frac{\int d\V{r}_\perp f^2(0,\V{r}_\perp)}{\int d\V{r}_\perp f(0,\V{r}_\perp)}=\frac{2}{3}\frac{g-g_{12}}{\pi R_r^2}  .
\end{equation}

\subsection{Experimental procedure}

In the experiment, we initially create a polarized Bose-Einstein condensate of $5 \times 10^5$ atoms in $|1,-1\rangle$. The atoms are trapped in an elongated optical trap \cite{Farolfi20} with radial and axial frequencies $\{\omega_x,\omega_r\}/2\pi=\{10(1),1006(1)\}$ Hz, with Thomas-Fermi radii $R_\mathrm{r}$ and $R_\mathrm{x}$ equal to \SI{2.2}{\um} and \SI{210}{\um}, respectively. 
A stable magnetic field of \SI{1.3018}{G} is applied along the $z$ axis to lift the degeneracy of the Zeeman substates. 
We stabilize the mixture against spin-changing collisions by lifting the energy of the $|1,0\rangle$ state through a microwave field blue-detuned respect to the $|1,0\rangle \rightarrow |2,0\rangle$ transition.
The high stability and uniformity of the field is ensured by the presence of a 4-layer magnetic shield \cite{Farolfi19} around the main vacuum cell, that suppresses the field fluctuations by more than 5 orders of magnitude down to 2 $\mu$G on the timescale of the full experimental sequence. At the used magnetic field, this corresponds to an energy fluctuation of about $\approx h\times \SI{3}{Hz}$. 
We suddenly switch on a microwave coupling between $|1,-1\rangle$ and $|1,+1\rangle$, by means of a two-photon transition, detuned by $\Delta$ from the intermediate state $|2,0\rangle$. The strength of the effective Rabi coupling $\Omega_\mathrm{R}$ is tuned by changing $\Delta$. 
The coupled states possess the peculiar feature that the intracomponent coupling constants $g_{-1}=g_{+1}=g$, while the interspecies one $g_{+1,-1}$ is $7\%$ smaller than $g$ \cite{Knoop2011}. This leads to miscibility \cite{Bienaime2016,Fava2018}. Our initial condition $\mathbf{s}= (0, 0, -n)$ corresponds, in the presence of Rabi coupling, to a highly excited state, far from the equilibrium point $\mathbf{s}= (n,0, 0)$. After a variable evolution time, we first image the density distribution of state $|1,-1\rangle$, then of state $|1,+1\rangle$. Both absorption pictures are taken after a short expansion time, respectively 2 ms and 3 ms after the switch off of coupling and trapping potential. We expand the sample in order to increase the signal to noise ratio of the measured magnetization, while the different expansion times are due to technical limitations. Density fluctuations in the region of strongly fluctuating magnetization are enhanced by the different expansion times.\\

In order to determine the one dimensional interaction parameter $\kappa$, we measured \cite{Farolfi21} the plasma oscillation frequency $\omega_p$, i.e., the frequency of the small magnetic fluctuations around the ground state $\mathbf{s}=(n,0,0)$. The latter can be indeed measured with high accuracy and can be directly compared with the simple analytical expression $\omega_p=\sqrt{\Omega_\mathrm{R}(\Omega_\mathrm{R}+\kappa n)}$. By measuring the plasma frequency in the central slice of the trap we extract $\kappa n_0$. The obtained value is in good agreement with the one resulting from the atom number and trapping frequencies. We checked that with such a value at hand we can reproduce the experimental spatial dependent plasma frequency simply by using the Thomas-Fermi one dimensional density $n(x)=n_0(1-x^2/R_\mathrm{x}^2)$.

\section{Acknowledgements}
We thank F. Dalfovo for his critical reading of the manuscipt and M. Oberthaler, G. Consolo, D. Go, E. Mendive-Tapia and N. Pavloff for fruitful discussions. 
We acknowledge funding from INFN through the FISH project, from the European Union’s Horizon 2020 Programme through the NAQUAS project of QuantERA ERA-NET Cofund in Quantum Technologies (Grant Agreement No. 731473) and from Provincia Autonoma di Trento. We thank the BEC Center, the Q@TN initiative and QuTip.

\section{Authors' contribution}
A.Recati and G.F. conceived the project.
A.F. performed the experiment.
A.F. and A.Z. analyzed the experimental data.
D.T. set up the experiment control.
A.G., A.Recati and A.Roy developed the theory and performed the corresponding numerical simulations. 
A.Recati, A.Z., and G.L. wrote the manuscript.
All authors contributed to the discussion and interpretation of the results.\\

\noindent $^*$ Correspondence to\\  
\href{mailto:giacomo.lamporesi@ino.cnr.it}{giacomo.lamporesi@ino.cnr.it}\\
\href{mailto:alessio.recati@ino.cnr.it}{alessio.recati@ino.cnr.it }\\
\href{mailto:alessandro.zenesini@ino.cnr.it}{alessandro.zenesini@ino.cnr.it}\\

\noindent  $^\dagger$ Current address: CNR Nanotec, Institute of Nanotechnology, via Monteroni, 73100, Lecce, Italy.

\noindent  $^\ddagger$ Current address: Institute for Quantum Electronics, ETH Z\"urich, Switzerland.
 
\section{Competing interests}
The authors declare no competing interests.

\section{Data and Code Availability}
The data that support the findings of this study are available from the corresponding author upon reasonable request.

\bibliography{bibliography.bib}

%\printbibliography

%\newrefsection

%\newpage

%\printbibliography

%\newpage

 \begin{figure*}
  \includegraphics[width=1.6\columnwidth]{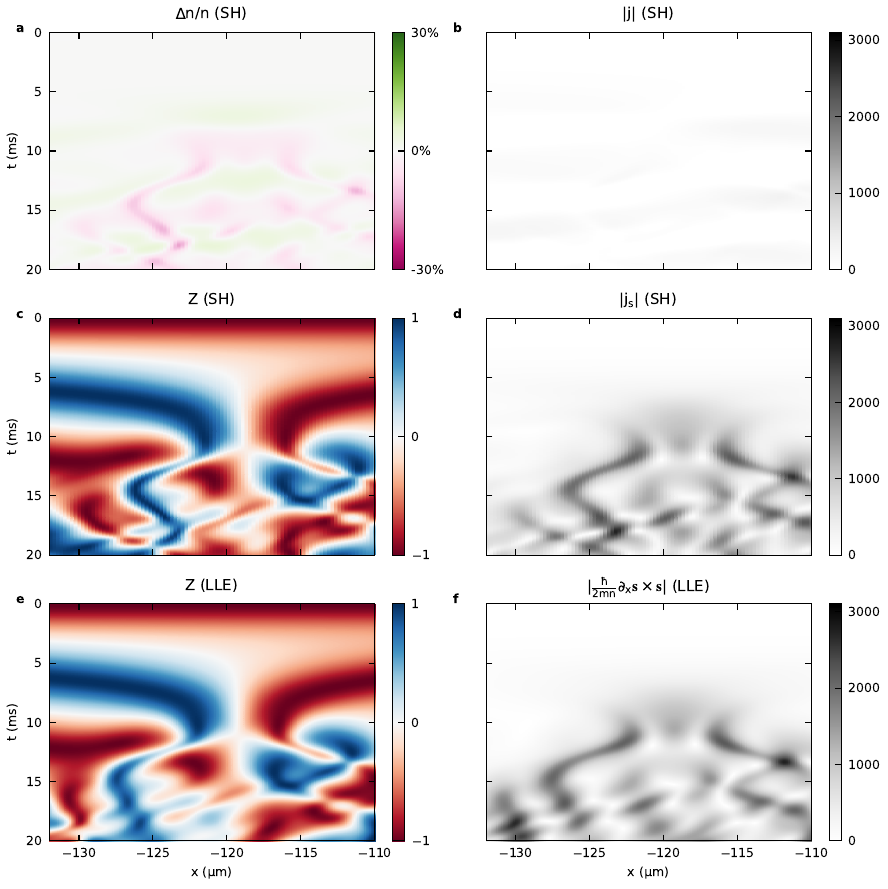}
  \caption{ \textbf{ExtendedDataFig.1  Comparison between spin hydrodynamic (SH) and LLE simulation.} Total density relative modulation (\textbf{a}), modulus of the total current (\textbf{b}), magnetization (\textbf{c}) and modulus of the spin current (\textbf{d}) simulated according to the SH around the interface before and after its breaking. The magnetization resulting from LLE (\textbf{e}) shows a very good agreement, accordingly to the negligible role of the density modulation in the dynamics. \textbf{f}, Contribution of the spin current term related to the quantum torque. Also in this case, panel \textbf{f} shows that the second term in Eq.~\ref{eq:j} dominates over the first one, hence the quantum torque drives the dynamics. The colorscale units of plots \textbf{b,d,f} is atoms/ms.}
  \label{fig:fig5}
\end{figure*}

\end{document}